# Evaluation of predictive correlation between flux expulsion and grain growth for superconducting radio frequency cavities

Zu Hawn Sung[1], Paulina Kulyavtsev[1,2], Martina Martinello[1,3], Dan Gonnella[3], Marc Ross[3], and Sam Posen[1]

[1] Applied Physics and Superconducting Technology Directorate, Fermi National Accelerator Laboratory, Batavia, IL, 60510, USA.

[2] Physics Department, Illinois Institute of Technology, Chicago, IL, 60616, USA

[3] SLAC National Accelerator Laboratory, Menlo Park, CA, 94025, USA

**Abstract:** A series of experiments were carried out in an effort to develop a simple method for predicting magnetic flux expulsion behavior of high purity niobium used to fabricate superconducting radio frequency (SRF) cavities. Using conventional metallographic characterizations in conjunction with high spatial resolution electron backscattered diffraction-orientation imaging microscopy (EBSD-OIM), we found that the flux expulsion behavior of 1.3 GHz single cell SRF Nb cavities is significantly associated with the grain growth of the Nb material during heat treatment. Most of Nb grains rapidly grew during 900 ºC heat treatment, and likely full-recrystallized with 1000 ºC HT. With comparison of the magnetic flux expulsion ratio ($B_{sc}/B_{nc}$) at $\Delta T$ = 5 K, the flux expulsion efficiency of the cavities increases along with increasing of grain size. Most interestingly, 900 ºC HT shows a roughly linear trend that suggests this criterion could be used to predict appropriate heat treatment temperature for sufficient flux expulsion behavior in SRF-grade Nb. This result would be used to see if flux expulsion can be predicted by examining the materials coming from the Nb vendor, prior to cavity fabrication.

## 1. Introduction

The transition metallic superconducting niobium has been widely accepted for superconducting RF resonators for linear particle accelerators, which are used to transfer energy to particle beams [1]. Its high ductility makes it possible to fabricate complicated RF cavity shapes and its excellent surface superconducting property enables intrinsic cavity quality factor ($Q_0$.) of

~$10^{10}$, leading to extremely efficient operation in particle accelerator applications [2]. Several different surface treatments in conjunction with thermal treatments are employed for maximizing the $Q_0$. In the nitrogen (N) doping process [3], Nb cavities are heat-treated at high temperatures in the presence of nitrogen to create nitrogen interstitials within the first few hundreds nanometer depth of the RF surface. This process, developed in 2013, makes it possible to improve quality factors by a factor of 2-3, compared to the previous state of the art technique [4,5]. N-doping has been implemented in the production of SRF Nb cavities for LCLS-II, an X-ray free electron laser (FEL) under construction at SLAC [6]. LCLS-II will operate in continuous wave (CW) mode, resulting in much larger heat dissipation by the cavities than if they were operated in pulsed mode. N-doping on the cavity to increase quality factor, therefore, enables a significant reduction in cryogenic plant infrastructure and operating cost. In addition to N-doping, other processes have been developed to modify the RF surface of cavities and increase quality factors, such as nitrogen infusion [7,8] and medium temperature baking [9].

While optimizing the RF surface superconductivity is crucial to achieving high $Q_0$, the bulk property is also of great importance. External magnetic flux can be trapped in the bulk of SRF Nb cavities during superconducting transition over the critical superconducting temperature of Nb, $T_c$ ~ 9.2 K, as a result, significant degradation of RF performance occurs [10,11]. Romanenko *et al.* [12] investigated the effect of thermal gradient (i.e. temperature difference ($\Delta T$) over the cavity surface) on flux expulsion, finding the cooling with a large thermal gradient promotes the expulsion of magnetic flux, while cooling slowly and uniformly tends to trap all ambient fields in the cavity wall. Posen *et al.* [13] investigated the effect of surface and bulk properties on the efficiency of magnetic flux trapping in single SRF Nb cavities, prepared by different cavity fabrication methods, finding that some cavities tend to trap most flux even when cooled with a large thermal gradient, but that high temperature heat treatment at 900 °C or above makes it possible for these cavities to achieve strong expulsion during cooldown.

Keeping trapped flux to an acceptable level was an important factor in achieving the $Q_0$ goals of LCLS-II cryomodules [14]. Magnetic shielding was a key contributor to this, but even after shielding, the specification for ambient magnetic fields was 5 mG (milli-gauss) at the cavity, and fully trapping this field would cause the cavities to fall below the $Q_0$ specification. Cooling with a high temperate gradient typically helps to expel flux, but the practical maximum gradient achievable during cooldown of a cryomodule is ~ 0.3 K/cm, corresponding to a $\Delta T$ ~ 5 K for a single cell cavity. For the as-received LCLS-II production material, near-full trapping would occur at this thermal gradient for a typical heat treatment at 800 °C for 3 hours.

One can try to infer possible types of defects that can admit magnetic flux based on correlations between cavity treatment and expulsion behavior. High temperature heat treatment improves flux expulsion behavior, surface treatment has a minimal effect, and mechanical deformation tends to result in strong trapping behavior [15–18]. These results suggest that intrinsic crystalline defects in Nb bulk like grain boundaries (GBs), tangled dislocations, low angle grain boundaries (LAGBs), or impurities like hydrides and nitrides, could be candidates for trapping sites of magnetic flux lines [19–24]. However, these defects must be comparable in size due to the relatively longer coherence length of superconducting niobium, $\xi_{Nb} \sim 50$ nm [25]. There have been efforts to associate defects with flux trapping in SRF-grade niobium by comparing microscopic studies of strongly and weakly trapping materials, but, to date, no specific defect have been conclusively identified. As such, it has been difficult to develop modified niobium production specifications to guarantee strong expulsion behavior. This produces challenges for projects such as the high energy upgrade to LCLS-II, LCLS II HE, for which strongly expelling materials must be obtained for cavity production. The solution implemented for LCLS-II was to buy typical material without modified specifications, and then to make single cell cavities from each heat treatment lot, and evaluate the temperature needed to achieve strong expulsion in the single cells in order to determine the treatment temperature for the production cavities made from the corresponding materials. However, such a process of fabricating and evaluating a series of single cell cavities is burdensome, and it is not practically applicable for the project like LCLS-II. Thus, the ultimate goal of our study is to bypass this process by a simpler method.

This study is based around a hypothesis of material defects for preferential pinning of magnetic flux: is it possible that the defects that trap flux during cooldown are the same defects that pin grain boundaries and determine grain growth during heat treatment? If so, we expect there to be a correlation between flux expulsion behavior and grain growth during heat treatment. The study has a number of advantages:
1. If any strong correlation is found, it could be easily implemented in the cavity fabrication process. Niobium vendors could measure grain growth in coupon samples from each heat treatment lot and sort the material according to the temperature at which sufficient grain growth was observed. Cavity vendors could manufacture cavities with the sorted materials, performing heat treatment to improve expulsion at the temperature indicated by the niobium vendor.
2. Measurement of grain size is a fairly standard process. Once this standard is established, niobium vendors can indicate specifications of the Nb materials for proper flux expulsion without additional measurements (e.g. cryogenic characterization).
3. The study can be carried out using existing resources. A number of cavities have already been characterized for flux expulsion after 900 °C for 3 hours heat treatment, and extra niobium material from the sheets used to fabricate these cavities is in hand. All that remains is to measure grain growth as a function of heat treatment.
4. The study is agnostic to the specific type of defect that causes pinning. It does not require detailed microscopic studies at the nm-scale.

Why should flux expulsion be correlated to grain growth? During heat treatment, Nb grains merge and grow to recover stored strain energy [26], and partial or full recrystallization would be achieved at a certain level (note that typical specifications require full or near-full recrystallization). However, this process is generally affected by pinning of grain boundaries on intrinsic defects or limited by dislocation mobility due to the defects, and as a result, potential flux pinning sites can form in the Nb bulk [16,17,27]. Some models suggested that defects that trap flux during cool down are reduced with high temperature heat treatment. As these defects are reduced, there would be a large mobility for grain boundaries during the heat treatment, allowing greater grain growth. High thermal energy also enhances dislocation movement, resulting in active coalition of grains. Thus, this study aims to classify grain growth behavior of SRF-Nb materials as a function of heat treatment temperature in order to account for magnetic flux expulsion properties of the SRF Nb cavity. From the study, a practical application would establish a method to predict flux expulsion behavior of the SRF Nb cavity from the starting material, prior to successive fabrication procedure.

## 2. Experimental

A series of 1.3 GHz single-cell SRF cavities made of various Nb sheets from the different vendors were characterized in magnetic flux expulsion behavior. All cavities were fabricated in the same way. Variation of external magnetic flux density (B) on the outside of the cavity wall was measured under the influence of magnetic field while cooling the cavity with cold helium gas flow applied parallel to the beam axis. The cold gas flow applied in this way generates a thermal gradient (ΔT) across the cavity during superconducting transition [12].

For bulk structural investigation, 10 mm × 5 mm × 2.5 mm ($l \times w \times t$) coupon samples were carefully extracted from the residual parts of the Nb sheets used to fabricate the single cell cavities by wire-EDM (electric discharge machining), and then thermally treated from 800 °C to 1000 °C for 3 hours at the same vacuum condition ($10^{-6}$ torr) in the same furnace of cavity treatment. After heat treatments, the coupons were cross-sectionally sliced with a precision diamond saw, and the opened surfaces were mechanically polished after mounting on a conductive puck, following the established polishing procedures [28]. The surfaces were finalized with 0.05 μm silica oxide particle solution within a vibratory polisher for 24-48 hours, in order to obtain defect and scratch-free surfaces. The list of the single cell cavities, SRF Nb sheets, and sample materials are tabulated in Table 1.

In this study, two different grain size evaluation methods were implemented; 1) conventional procedures with optical imaging analysis and 2) automated evaluation by backscattered electron diffraction – orientation imaging microscopy (EBSD-OIM), in order to verify compatibility of the two procedures. The conventional method was employed based on ASTM E-112[13] [29] after etching the polished cross-sectional surfaces of the samples with buffered chemical polishing (BCP; 1:1:2 volumetric ratio of the mixture of HF: $HNO_3$: $H_2SO_4$) for 8 min. BCP was chosen because it tends to reveal grain boundary (~ 1-100 μm depth) [2], helping to reconstruct grain structure. Using scanning laser confocal microscope (SLCM), we were able to resolve such non-uniform GB topology. SLCM is superior to conventional optical microscope in Z-resolution (normally surface roughness) due to its deep focal depth resolution [30]. The grain structure of the Nb materials was further characterized with backscattered electron diffraction – orientation imaging microscopy (EBSD-OIM), equipped in JEOL JSM-5900LV scanning electron microscope. The extensive area (~ 2 $mm^2$) of the polished cross-section surfaces were scanned, and the grain sizes and GB morphology were characterized with Oxford HKL Channel 5 software after smoothing un-indexed pixels with 0.5-1 degree threshold to minimize interference of residual scratches on automated calculation.

## 3. Results

Figure 1 compares magnetic flux expulsion ratios ($B_{sc}/B_{nc}$) of the single cell cavities as a function of thermal gradient ($\Delta T$). $B_{sc}/B_{nc}$ is defined as the ratio of the B at the normal conducting state to superconducting state, below the critical superconducting temperature of niobium, $T_c \sim 9.2$ K. $B_{sc}/B_{nc} \approx 1.7$ corresponds to the cavity completely expelling external magnetic flux during superconducting cooling. At $B_{sc}/B_{nc} \approx 1$, most of magnetic flux is trapped in the walls of the cavity. Each cavity shows different expulsion behavior even having the same heat treatment history, 900 °C for 3 hrs at high (10-6 torr) vacuum, as well as cell fabrication. The cavity RDT-NX02 traps most of the external magnetic flux even at relatively large thermal gradient across the cavity ($\Delta T \sim 5$ K and even up to ~15 K). In contrast, cavity SC02 shows extremely strong expulsion behavior, reaching near-full expulsion with a $\Delta T$ of just 2 K.

Among several procedures of grain size evaluation with optical microscopy presented in ASTM E-112[13] [29], the planimetric (Jeffries) general intercept method was mainly used in this study. The planimetric procedure was easily applicable to SRF Nb materials without limitations on imaging size and magnification, and the results were comparable to EBSD-OIM analysis. We also introduce the Abrams three-circles circular intercept procedure to clarify efficiency of conventional grain size evaluations. As a first step, we characterized grain size of the A4 (TE1AES024) coupon after 800 °C/3 hrs heat treatment. Figure 2 shows a laser intensity image of BCP'ed grain boundaries from scanning laser confocal microscopy (SLCM). The image was acquired at 100X magnification with a size of 2048 × 1536 pixels using 10X objective lens. The marked areas with A, B, and C in Fig. 2a have different roughness at GBs, $R_z$ = 7.09 μm, 3.71 μm, and 1.33 μm, respectively. The planimetric (Jeffries) procedure is defined with an arbitrary circle containing ~ 40-50 grains (Fig. 2a). By counting inside and intercepted grains with the red circle ($N_{inside}$ = 69 and $N_{intercepted}$ = 31) and applying Jeffries' multiplier ($f$ = 2 at 100X magnification), grains per square millimeter ($\overline{N_A}$) of the A4 is obtained as 169 from $N_A = f(N_{inside} + N_{intercepted}/2)$, which is referred to 4.0-4.5 of grain size number ($G$), describing grain diameter between 75.5 and 89.8 μm according to ASTM E-112[13]. Table 2 is a part of the table for grain size relationships computed for uniform, randomly oriented, equiaxed grains, in ASTM E112[13]. ASTM $G$ number decreases with increasing of density of grains and grain size. For Abrams' circular intercept procedure, three circles were drawn in the laser intensity image (Fig. 2b) after proportionally reduced the total circumference from 500 mm (250 mm, 166.7 mm, and

83.3 mm) to 5.3 mm (2.52 mm, 1.69 mm, and 1.08 mm); our image size (2048 × 1536 pixels: 1.45 mm × 1.09 mm at 100X) was limited in a space, compared to Abrams's one (100 mm × 100 mm at 1X mag). Similarly, by counting total number of intercepts ($N_i$) of the three-blue circles with GBs and calculating the number of intercepts counted on the field ($\overline{N_L} = N_i/(L/M)$) (the total test length: $L, = 500$ mm, magnification: $M, = 100X$, and the linear intercept value: $\bar{l}, = (1/\overline{N_L})$), grain size number ($G$) is characterized in 3.9 according to the formulae provided; $G = -3.2877 - 6.6439 \log_{10} \bar{l}$, $G = -3.2877 + 2\log_2 \overline{N_L}$, or $G = -3.2877 + 6.6439 \log_{10} \overline{N_L}$ (A1.4, 1.5, and 1.6 equations in A1 annexes of ASTM E-112[13]), which gives $\overline{N_A}$ = 87.68-124 grains/mm$^2$ and $\bar{d}$=89.8-106.8 µm, respectively. Abrams' circular procedure shows slightly bigger grain size than the planimetric procedure does, but this three-circular intercept method was no longer applicable for above 950 °C HT due to limited space by significant grain growth.

Grain structure of the 800 °C HT'ed A4 sample was also characterized with EBSD-OIM. Figure 3 shows grain orientation and GB morphology of the finely polished cross-sectional surface. Although several scratches remained from mechanical polishing, as shown in the brightness and contrast image of kikuchi pattern intensities (Fig. 3a), grain diameter, grains numbers, and grain locations were evaluated within the field of scanning view based on inverse pole figure (IPF) by HKL Channel 5 software. Figure 4 illustrates size distributions of inner and edge or corner grains of the 800 °C HT'ed A4 sample, presenting total 263 grains on the cross-section (1.16 mm$^2$). The average grains per square millimeter and grain size, (282 grains/mm$^2$, 65.2 µm) are found to be $G$ ~ 5.0-5.5 and 4.5-5.0, respectively, showing ~ 15 % smaller than those obtained from both conventional methods. Figure 5 illustrates IPF and grain size distribution profiles of the S6 (cavity SC06) as heat treatment temperature increases. Rapid grain growth is seen after 900 °C HT, and huge recrystallization occurred during 1000 °C HT. Frequency (counts) of edge and corner grains outnumbered those of the inner grains as Nb grain growth continues, as a result, the blue histogram line passes over the orange line after 1000 °C HT.

The two conventional procedures provided slightly different but overall consistent grain sizes. However, only the planimetric (Jeffries') procedure was applied for the entire range of heat treatment temperatures because extended grain growth during 950 °C/3 hrs HT restricted the applicability of the Abram's procedure. Table 3 describes ASTM grain size number ($G$) of the SRF

Nb materials along with HT temperature, defined by Jeffries' procedure. Most of the Nb samples experienced rapid grain growth during 900 °C/3 hrs HT, resulting in significant reduction of ($G$), but N2 (NX02) showed the least amount of grain growth, even after 975 °C HT. This trend is similarly observed with EBSD-OIM study. Figure 6 shows variation of the average grain sizes of the Nb samples with heat treatment temperature, characterized by EBSD-OIM. Grain sizes of most of the Nb sample are doubled after 900 °C HT. Likewise, the N2 shows the least amount of grain growth even after 1000 °C HT. ASTM Grain size No ($G$) based on the EBSD-OIM analysis is listed in Table 4. N2 and S2 have the smallest and largest grain size, respectively, just after 800 °C HT. After 900 °C HT, $G$ of most samples decrease in half except N1, N2 and S8. However, S8 shows steady grain growth with HT temperature.

Magnetic flux expulsion ratios ($B_{sc}/B_{nc}$) of the single cavities at $\Delta T$ = 2 K and 5 K are plotted with the average grain sizes of the Nb coupon materials after 900 °C/3 hrs HT (Figure 7), which is determined by EBSD-OIM study. Large thermal gradient $\Delta T$ = 5 K provides better flux expulsion ratios with the same grain size, compared to the cases at $\Delta T$ = 2 K. Cavity SC02 (S2) shows outstanding expulsion behavior close to $B_{sc}/B_{nc} \approx 1.7$ at both cooling gradients ($\Delta T$) after heat treatment at 900 °C. Cavity AES024 (A4) shows similar optimum flux expulsion with somewhat smaller grain size while it has slightly lower $B_{sc}/B_{nc} < 1.7$ at $\Delta T$ = 2 K. Flux expulsion ratios of the cavities are mostly enhanced at $\Delta T$ = 5 K with increasing of grain size. This improvement is suppressed with $\Delta T$ = 2 K. However, the NX02 cavity does not show any change with grain growth. Interestingly, a monotonically increasing trend between $B_{sc}/B_{nc}$ and grain size at $\Delta T$ = 5 K after 900 °C/3 hrs HT, is observed, as guided by the red-dotted line in Fig. 7b). This behavior suggests that there is a criteria of heat treatment temperature to effectively improve flux expulsion behavior of the cavities in some SRF Nb materials.

**Discussion:**

Efficiency of external magnetic expulsion during the superconducting transition is a key contributor to optimizing the RF performance of SRF Nb cavities [12–14]. It is already known that the flux expulsion ratio ($B_{sc}/B_{nc}$) is significantly related with bulk structural property rather than surface characteristics of the cavity, determined by chemical or low temperature treatments or both combined treatments, i.e. nitrogen gas treatment and 120-300 °C mild heat treatment [3,7–9].

Niobium is a marginal type II superconductor, having a narrow range of the mixed state with long superconducting coherence length, $\xi_{Nb}$~50 nm, compared to other composite type II superconductors [25]. Thus, the size of crystalline structural defects for preferential trapping of magnetic flux lines should be comparable to $\xi_{Nb}$ ~ 50 nm. Grain boundaries [24,31], tangled dislocations [16,22], low angle (< 2°) grain boundaries [17], dense of dislocations networks, or impurities like nitride, carbide, or hydride [20,32,33] would be strong candidates for trapping sites if their size are compromised.

Recrystallization and/or grain growth typically occur during SRF Nb cavity thermal treatment, though the extent to which they occur can vary. Impurities locally pinned at bulk Nb lattice can deteriorate recrystallization or grain growth by interfering movement of grain boundaries [26]. Annihilation of dislocations by climbing, gliding, or cross slipping resulting from applied thermal energy is another important aspect for grain growth, but the procedure highly relies on initially stored strain energy from cold work deformation [34,35]. The impurities also disturb the dynamics of dislocation movement. Therefore, non-fully completed recrystallization may result in preferential flux trapping sites such as tangled dislocations or low angle (< 2°) grain boundaries (LAGBs) [36]. Investigation for the effect of ppm level impurities on grain growth behavior is extremely challenging, especially for the SRF-grade Nb [37]. It is also challenging to evaluate the effect of dislocation movement on grain growth for large area of Nb bulk structure without information about stored strained energy from cold work deformation. Given these restrictions, quantifying of Nb grain size with a series of heat treatments in the same fashion of cavity treatment would be an alternative way to understand the extent of which the defects influence on recrystallization, which is attributed to variation of magnetic flux trapping sites in the bulk of the SRF Nb cavity. Furthermore, variations can occur in fabrication processes for SRF-grade Nb sheets, so that recrystallization and recovery process could be different, resulting in variation of flux expulsion ratio ($B_{sc}/B_{nc}$) with heat treatment temperature. Therefore, the goal of our study is to establish a practical way if flux expulsion can be gauged with bulk structural analysis on a starting material, prior to cavity fabrication. Studying sample should be prepared at the same condition of cavity fabrication.

A flux expulsion ratio of $B_{sc}/B_{nc} \approx 1.7$ corresponds to complete or near complete expulsion of external magnetic flux. Only SC02 and TE1AES024 cavities meet this requirement with $\Delta T \leq$ 4 K with 900 °C HT; some other cavities achieve close to this ratio with $\Delta T \geq 5$ K, as shown in

Fig. 1. However, few cavities still have poor flux expulsion ratio regardless of cooling gradient. In particular, the cavity NX02 trapped nearly all fluxes during superconducting transition, even for large ΔT. The N2 materials used for the NX01 cavity also shows by far the least amount of grain growth even after 1000 °C HT.

The bulk structural properties of the SRF Nb materials were evaluated in terms of ASTM grain size $G$ based on Table 2 for grain size relationships computed for uniform, randomly oriented, equiaxed grains from [29]. Planimetric (Jeffries') and Abrams's circular intercept procedures showed similar grain size, but the sizes are ~10-15 % bigger than those from EBSD-OIM study. EBSD-OIM analysis is further compared to both conventional methods because EBSD-OIM discriminates grains based on crystal lattice orientation and identifies low angle GBs (< 1-2°) that are not clearly revealed with chemical etching like BCP due to minimal difference in surface energy between grains. However, EBSD-OIM is limited with both of scan area and scan step size, so that we compensated large scan area (1.8-2 mm$^2$) with proper scan step size (4-5 μm) in order to cover the entire cross section of the sample (2 mm × 2 mm) for better comparison. Nevertheless, this step size was not enough to recognize dislocation-related low angle grain boundaries like GNB (geometrical necessary grain boundaries), which are generally produced by mechanical deformation [38].

During 900 ºC/3 hrs HT, most of SRF Nb materials experienced rapid grain growth, and then the Nb grains gradually grew as heat treatment temperature increased up to 1000 ºC. S6 (materials from the cavity SC06) had grain diameters of half a millimeter and larger (Fig. 5), corresponding to $G$ ~ 00 after 1000 °C HT. In contrast, N2 (cavity NX-02) showed the least amount of grain growth, less than 100 μm of grain diameter, even after 1000 °C HT. This trend is well described in Table 6 (from Jeffries's procedure), and Table 4 and Figure 6 (from EBSD-OIM analysis). After 900 °C HT, ASTM grain size No. ($G$) of the all samples (except N2) significantly decreased as a result of ~35-50 % increasing of grain diameter, and then reaching close to $G$ ~ 00-0.0 with 1000 °C HT. S6 sample shows the highest grain growth rate after 1000 °C HT, but the SC06 cavity has $B_{sc}/B_{nc}$ ~ 1.57 at ΔT = 5 K with 900 °C HT, which is lower than the ratio of SC02, SC04, and TE1AES024 cavities ($B_{sc}/B_{nc}$ ~ 1.69, 1.59, and 1.68 respectively). Cavity SC02 has the highest $B_{sc}/B_{nc}$ at ΔT = 2 K and 5 K with 900 °C HT temperature, with the largest grain size ~200 μm. This result suggests that grain growth rate indicates the extent of which heat treatment remove defects in the bulk SRF Nb cavity. Thus, like S6 Nb materials, if heat treatment is applied above 1000 °C HT, flux expulsion behavior would be expected to further improve from 900 °C HT. However, such high temperature HT may also induce significant degradation on mechanical sustainability, so-called stiffness [39,40]. If the Nb cavity wall is too soft, it may result in undesirable detuning during handling [13,41]. This is why it is important to identify heat treatment temperatures that are sufficiently high to achieve strong flux expulsion without being excessively high.

Figure 7b) provides a guideline of how this type of analysis could be used to evaluate if a given heat treatment is sufficient to provide strong expulsion. Based on the requirements chosen for $B_{sc}/B_{nc}$ at a given $\Delta T$, one could set a minimum grain size for a given heat treatment. For example, the ~ 45 μm grain size of N2 may be considered too small (expected to result in poor expulsion) and therefore a higher heat treatment temperature would be selected where more grain growth was observed (or else possible rejected if insufficient growth was observed even in samples heat treated up to 1000 °C). By observing grain growth vs temperature for a series of samples from a given heat production lot, a heat treatment temperature and time could be recommended for that lot to the cavity vendor. For example, for S2 and A4, 900 °C for 3 hours seems to be sufficient. For A2, S4, and S8, a somewhat higher temperature ~ 925 °C or 950 °C for 3 hours may be recommended, where more growth was observed without becoming too high. This method would avoid cryogenic testing to evaluate flux expulsion. The vendor would not need high resolution grain size evaluation with EBSD-OIM because this study shows that simple planimetric circular intercept method is enough to measure grain size with expecting error percentage. This method for choosing heat treatment temperature has not yet been evaluated in production, but it would be interesting to test it, possibly in conjunction with at least some cryogenic testing for verification in a future production. We also evaluated the cavity TD01 (Nb sheet #: T10971, shown in Fig. 1) and its coupon samples, but the magnetic expulsion ratio and grain growth were inconsistent with the sheet locations, so that we decided to exclude the result from this paper. There might be inhomogeneity in the sheet materials. As Palczewski [42] pointed out, flux expulsion behavior of the cavities heated treated at the same temperature lot can vary with the batches from the same Nb vendor. However, our re-evaluation on another area of the TD1 coupon followed the similar trend guided by the red-dotted line with 152.7 μm of a grain size.

It is not known at this point how changes in the manufacturing process would change the grain size increase corresponding to the onset of strong flux expulsion. It would be interesting to study the effect on both flux expulsion and grain growth for changing sheet production parameters including purity, rolling procedure, annealing temperature/time, and roller levelling (or "kiss pass" rolling) after annealing. However, this would be a much more involved study. It would be also be interesting to look at the flux expulsion at different temperatures (beyond just 900 °C) in these cavities to see how well the grain growth at these temperatures corresponded to flux expulsion behavior at the same temperature.

Specific types of the defects responsible for preferential flux trapping on the SRF cavity could not be identified in this study, thus high-resolution microcopy will be performed in the future work. However, since a correlation between flux pinning sites and pinning sites for grain growth (assuming a pinning mechanism for both of these phenomena) were observed, it does hint at certain types of features for pinning flux, such as LAGBs and dislocation clusters, which may also pin grain growth. This add some additional information for identifying the underlying mechanism for variation in flux expulsion in different SRF-grade niobium.

In summary, we investigated the relationship between external magnetic flux expulsion behavior of the SRF Nb cavities and grain growth using various Nb materials by applying heat treatment temperature from 800 °C to 1000 °C for 3 hours. For quantitative analysis of Nb grain growth, two different characterization methods were compared; 1) conventional optical imaging analysis and 2) EBSD-OIM methods, in order to verify compatibilities of each methods. Variations of grain size along with heat treatment temperature are represented based on ASTM grain size No $G$. SRF Nb grains rapidly grew with 900 °C HT and mostly recrystallized after 1000 °C HT. The magnetic flux expulsion ratio ($B_{sc}/B_{nc}$) improved with increasing of grain size, but a rough and monochromatically increasing trend was observed at thermal gradient, $\Delta T = 5$ K, after 900 °C HT. This suggests that controlling grain size with heat treatment temperature would be the pathway of achieving optimum expulsion ratio in SRF Nb cavity, prior to practical cavity fabrication. This study is aiming not for fundamental investigation on recovery, recrystallization, and grain growth of SRF Nb materials, but for improving the understanding of magnetic flux expulsion behavior of the Nb cavity with starting materials. It is also strongly believed that the method that we applied for conventional grain size measurement would be a good reference to SRF Nb material and cavity vendor and SRF community.

**Acknowledgement:**

This work was supported by the United States Department of Energy, Offices of High Energy Physics and Basic Energy Science under Contract No. DE-AC05-06OR23177 (Fermi National Accelerator Laboratory) and DE-AC02-76F00525 (SLAC National Accelerator Laboratory). The authors would like to thank for technical assistance from the FNAL cavity preparation and cryogenic teams.

**Reference:**

[1] H. Padamsee, *50 Years of Success for SRF Accelerators—a Review*, Supercond. Sci. Technol. **30**, 053003 (2017).

[2] H. Padamsee, *RF Superconductivity: Science, Technology, and Applications*, 1 edition (Wiley-VCH, Weinheim, 2009).

[3] A. Grassellino, A. Romanenko, D. Sergatskov, O. Melnychuk, Y. Trenikhina, A. Crawford, A. Rowe, M. Wong, T. Khabiboulline, and F. Barkov, *Nitrogen and Argon Doping of Niobium for Superconducting Radio Frequency Cavities: A Pathway to Highly Efficient Accelerating Structures*, Superconductor Science and Technology **26**, 102001 (2013).

[4] J. P. Charrier, B. Coadou, and B. Visentin, *Improvements of Superconducting Cavity Performances at High Accelerating Gradients*, In Proceedings of the EPAC'98 1885 (1998).

[5] G. Ciovati, *Effect of low-temperature baking on the radio-frequency properties of niobium superconducting cavities for particle accelerator,* J of Appl. Phys, 96. 1591 (2004).

[6] D. Gonnella, S. Aderhold, A. Burrill, E. Daly, K. Davis, A. Grassellino, C. Grimm, T. Khabiboulline, F. Marhauser, O. Melnychuk, A. Palczewski, S. Posen, M. Ross, D. Sergatskov, A. Sukhanov, Y. Trenikhina, and K. M. Wilson, *Industrialization of the Nitrogen-Doping Preparation for SRF Cavities for LCLS-II*, Nuclear Instruments and Methods in Physics Research Section A: Accelerators, Spectrometers, Detectors and Associated Equipment **883**, 143 (2018).

[7] A. Grassellino, A. Romanenko, Y. Trenikhina, M. Checchin, M. Martinello, O. S. Melnychuk, S. Chandrasekaran, D. A. Sergatskov, S. Posen, A. C. Crawford, S. Aderhold, and D. Bice, *Unprecedented Quality Factors at Accelerating Gradients up to 45 MVm$^{-1}$ in Niobium Superconducting Resonators via Low Temperature Nitrogen Infusion*, Superconductor Science and Technology **30**, 094004 (2017).

[8] P. Dhakal, S. Chetri, S. Balachandran, P. J. Lee, and G. Ciovati, *Effect of Low Temperature Baking in Nitrogen on the Performance of a Niobium Superconducting Radio Frequency Cavity*, Phys. Rev. Accel. Beams **21**, 032001 (2018).

[9] S. Posen, A. Romanenko, A. Grassellino, O. S. Melnychuk, and D. A. Sergatskov, *Ultralow Surface Resistance via Vacuum Heat Treatment of Superconducting Radio-Frequency Cavities*, Phys. Rev. Applied **13**, 014024 (2020).

[10] A. Gurevich and G. Ciovati, *Dynamics of Vortex Penetration, Jumpwise Instabilities, and Nonlinear Surface Resistance of Type-II Superconductors in Strong Rf Fields*, Phys. Rev. B **77**, 104501 (2008).

[11] A. Gurevich, *Multiscale Mechanisms of SRF Breakdown*, Physica C: Superconductivity **441**, 38 (2006).

[12] A. Romanenko, A. Grassellino, O. Melnychuk, and D. A. Sergatskov, *Dependence of the Residual Surface Resistance of Superconducting Radio Frequency Cavities on the Cooling Dynamics around Tc*, Journal of Applied Physics **115**, 184903 (2014).

[13] S. Posen, M. Checchin, A. C. Crawford, A. Grassellino, M. Martinello, O. S. Melnychuk, A. Romanenko, D. A. Sergatskov, and Y. Trenikhina, *Efficient Expulsion of Magnetic Flux in Superconducting Radiofrequency Cavities for High $Q_0$ Applications*, Journal of Applied Physics **119**, 213903 (2016).

[14] S. Posen, G. Wu, A. Grassellino, E. Harms, O. S. Melnychuk, D. A. Sergatskov, N. Solyak, A. Romanenko, A. Palczewski, D. Gonnella, and T. Peterson, *Role of Magnetic Flux Expulsion to Reach $Q_0 > 3 \times 10^{10}$ in Superconducting Rf Cryomodules*, Phys. Rev. Accel. Beams **22**, 032001 (2019).

[15] S. Posen, S. K. Chandrasekaran, A. C. Crawford, A. Grassellino, O. Melnychuk, A. Romanenko, and Y. Trenikhina, *The Effect of Mechanical Cold Work on the Magnetic Flux Expulsion of Niobium*, arXiv:1804.07207.

[16] C. Z. Antoine, *Influence of Crystalline Structure on RF Dissipation in Niobium*, Phys. Rev. Accel. Beam 22, 034801, (2019).

[17] Z.-H. Sung, M. Wang, A. A. Polyanskii, C. Santosh, S. Balachandran, C. Compton, D. C. Larbalestier, T. R. Bieler, and P. J. Lee, *Development of Low Angle Grain Boundaries in Lightly Deformed Superconducting Niobium and Their Influence on Hydride Distribution and Flux Perturbation*, Journal of Applied Physics **121**, 193903 (2017).

[18] Z.-H. Sung, A. Dzyuba, P. J. Lee, D. C. Larbalestier, and L. D. Cooley, *Evidence of Incomplete Annealing at 800 °C and the Effects of 120 °C Baking on the Crystal Orientation and the Surface Superconducting Properties of Cold-Worked and Chemically Polished Nb*, Supercond. Sci. Technol. **28**, 075003 (2015).

[19] A. Romanenko, F. Barkov, L. D. Cooley, and A. Grassellino, *Proximity Breakdown of Hydrides in Superconducting Niobium Cavities*, Superconductor Science and Technology **26**, 035003 (2013).

[20] F. Barkov, A. Romanenko, Y. Trenikhina, and A. Grassellino, *Precipitation of Hydrides in High Purity Niobium after Different Treatments*, Journal of Applied Physics **114**, 164904 (2013).

[21] A. Romanenko and L. V. Goncharova, *Elastic Recoil Detection Studies of Near-Surface Hydrogen in Cavity-Grade Niobium*, Supercond. Sci. Technol. **24**, 105017 (2011).

[22] M. Wang, D. Kang, and T. R. Bieler, *Direct Observation of Dislocation Structure Evolution in SRF Cavity Niobium Using Electron Channeling Contrast Imaging*, Journal of Applied Physics **124**, 155105 (2018).

[23] Z.-H. Sung, *Direct Correlation of State of the Art Cavity Performance With Surface Nb Nano-Hydrides Cutouts Observed via Cryogenic AFM*, (unpublished).

[24] Z.-H. Sung, P. J. Lee, A. Gurevich, and D. C. Larbalestier, *Evidence for Preferential Flux Flow at the Grain Boundaries of Superconducting RF-Quality Niobium*, Supercond. Sci. Technol. **31**, 045001 (2018).

[25] M. Tinkham, *Introduction to Superconductivity: Second Edition*, Second edition (Dover Publications, Mineola, NY, 2004).

[26] R. Abbaschian and R. E. Reed-Hill, *Physical Metallurgy Principles*, 4 edition (Cengage Learning, Stamford, CT, 2008).

[27] T. R. Bieler, D. Kang, D. C. Baars, S. Chandrasekaran, A. Mapar, G. Ciovati, N. T. Wright, F. Pourboghrat, J. E. Murphy, C. C. Compton, and G. R. Myneni, *Deformation Mechanisms,*

*Defects, Heat Treatment, and Thermal Conductivity in Large Grain Niobium*, AIP Conference Proceedings **1687**, 020002 (2015).

[28] Z.-H. Sung, P. J. Lee, and D. C. Larbalestier, *Observation of the Microstructure of Grain Boundary Oxides in Superconducting RF-Quality Niobium With High-Resolution TEM (Transmission Electron Microscope)*, IEEE Transactions on Applied Superconductivity **24**, 68 (2014).

[29] E04 Committee, Test Methods for Determining Average Grain Size, ASTM International, n.d.

[30] A. Dzyuba, A. Romanenko, and L. D. Cooley, *Model for Initiation of Quality Factor Degradation at High Accelerating Fields in Superconducting Radio-Frequency Cavities*, Supercond. Sci. Technol. **23**, 125011 (2010).

[31] A. Dasgupta, C. C. Koch, D. M. Kroeger, and Y. T. Chou, *Flux Pinning by Grain Boundaries in Niobium Bicrystals*, Philosophical Magazine B **38**, 367 (1978).

[32] Y. Trenikhina, A. Romanenko, J. Kwon, J.-M. Zuo, and J. F. Zasadzinski, *Nanostructural Features Degrading the Performance of Superconducting Radio Frequency Niobium Cavities Revealed by Transmission Electron Microscopy and Electron Energy Loss Spectroscopy*, Journal of Applied Physics **117**, 154507 (2015).

[33] R. D. Veit, R. G. Farber, N. S. Sitaraman, T. A. Arias, and S. J. Sibener, *Suppression of Nano-Hydride Growth on Nb(100) Due to Nitrogen Doping*, J. Chem. Phys. **152**, 214703 (2020).

[34] T. R. Bieler, N. T. Wright, F. Pourboghrat, C. Compton, K. T. Hartwig, D. Baars, A. Zamiri, S. Chandrasekaran, P. Darbandi, H. Jiang, E. Skoug, S. Balachandran, G. E. Ice, and W. Liu, *Physical and Mechanical Metallurgy of High Purity Nb for Accelerator Cavities*, Phys. Rev. ST Accel. Beams **13**, 031002 (2010).

[35] S. Balachandran, Microstructure Development in Bulk Niobium Following Severe Plastic Deformation and Annealing. Doctoral Dissertation, Texas A&M University, 2015.

[36] S. Balachandran, A. Polyanskii, S. Chetri, P. Dhakal, Y.-F. Su, Z.-H. Sung, and P. J. Lee, *Direct Evidence of Microstructure Dependence of Magnetic Flux Trapping in Niobium*, Sci Rep **11**, 5364 (2021).

[37] W. Singer, X. Singer, A. Brinkmann, J. Iversen, A. Matheisen, A. Navitski, Y. Tamashevich, P. Michelato, and L. Monaco, *Superconducting Cavity Material for the European XFEL*, Supercond. Sci. Technol. **28**, 085014 (2015).

[38] D. A. Hughes, N. Hansen, and D. J. Bammann, *Geometrically Necessary Boundaries, Incidental Dislocation Boundaries and Geometrically Necessary Dislocations*, Scripta Materialia **48**, 147 (2003).

[39] Y. Yamaguchi, H. Doryo, M. Yuasa, H. Miyamoto, and M. Yamanaka, *Deformation and Recrystallization Behavior of Super High-Purity Niobium for SRF Cavity*, IOP Conf. Ser.: Mater. Sci. Eng. **194**, 012029 (2017).

[40] G. R. Myneni, *Physical and Mechanical Properties of Niobium for SRF Science and Technology*, AIP Conference Proceedings **927**, 41 (2007).

[41] A. Arnold, P. Murcek, J. Teichert, and R. Xiang, *Fabrication, Tuning, Treatment and Testing of Two 3.5 Cell Photo-Injector Cavities for the ELBE Linac*, Proceedings of SRF2011, Chicago, IL, USA **TUPO019**.

[42] A. Palczewski, D. Gonnella, O. Melnychuk, and D. Sergatskov, *Study of Flux Trapping Variability between Batches of Tokyo Denkai Niobium Used for the LCLS-II Project and Subsequent 9-Cell RF Loss Distribution between the Batches*, Proceedings of the 19th International Conference on RF Superconductivity **SRF2019**, 6 pages (2019).

Table 1. The list of the single cell SRF Nb cavities, SRF-grade Nb sheets, and coupon ID, studied in this paper.

| Serial # of SRF grade Nb sheet | Cavity # | Coupon ID |
|---|---|---|
| T10971 | TD02 | T2 |
| N10786 | NX01 | N1 |
| N11302 | NX02 | N2 |
| T13688 | SC04 | S4 |
| T13971 | SC06 | S6 |
| T13128 | SC08 | S8 |
| T13456 | SC02 | S2 |
| 250 | TE1AES024 | A4 |
| 78XX | TE1AES022 | A2 |

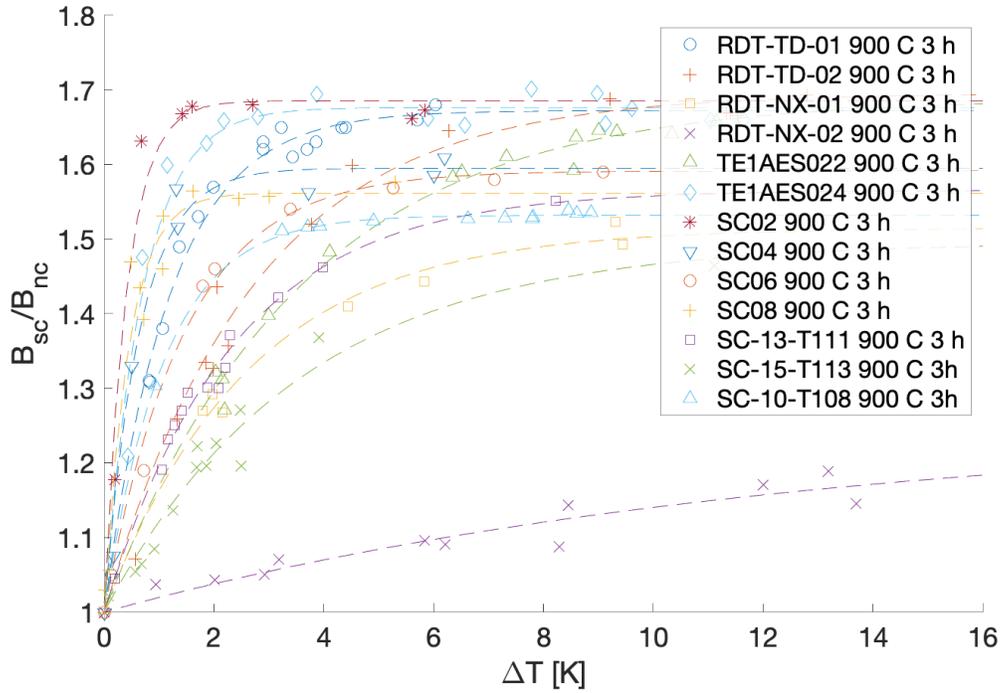

FIG. 1. Magnetic flux expulsion rate ($B_{sc}/B_{nc}$) of the single cell SRF Nb cavities depending on cooling gradient ($\Delta T$ [K]), after 900 °C/ 3 hrs heat treatment at $10^{-6}$ torr vacuum.

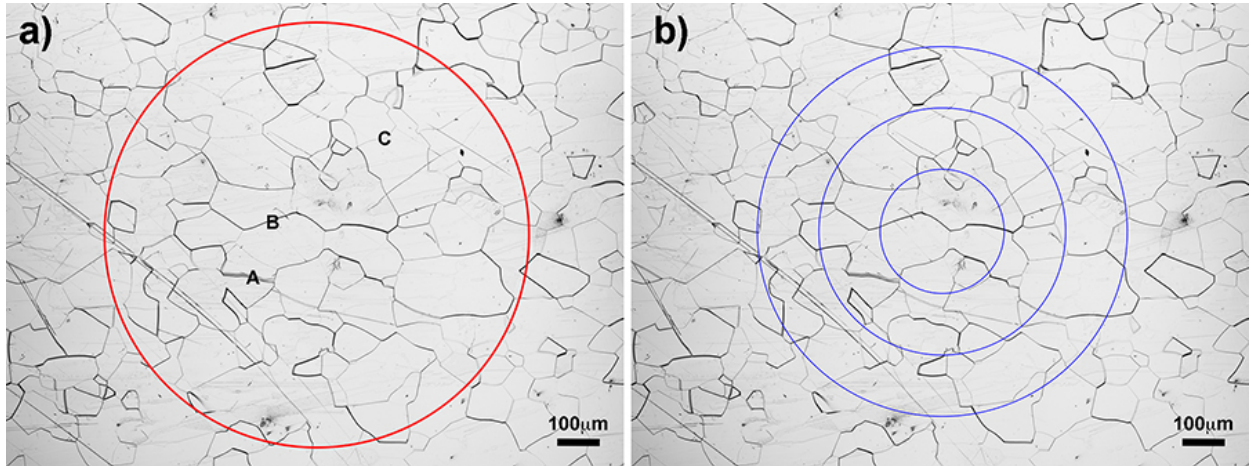

FIG. 2. Laser intensity surface image (2048 × 1536 pixels) of BCP'ed (chemically etched) A4 (TE1AES024) after 800 °C/3 hrs heat treatment, at 100X magnification by scanning laser confocal microscope (SLCM). a) Planimetric (Jeffries) procedure with an arbitrary circle (red) and b) Abrams circular intercept procedure with three circles (blue). A, B, and C represent the areas having different roughness at grain boundaries (GBs)

Table 2. Grain size relationships computed for uniform, randomly oriented, equiaxed grains, a part of Table 4 in ASTM E112-13 [29]. Reproduction with copyright certified from International ASTM

| Grain Size No. [G] | $\overline{N_A}$ (Grains/Unit Area) | $\bar{d}$ (Average Diameter) |
|---|---|---|
| | No./mm$_2$ at 1× | μm |
| 00 | 3.88 | 508.0 |
| 0 | 7.75 | 359.2 |
| 0.5 | 10.96 | 302.1 |
| 1.0 | 15.50 | 254.0 |
| 1.5 | 21.92 | 213.6 |
| 2.0 | 31.00 | 179.6 |
| 2.5 | 43.84 | 151.0 |
| 3.0 | 62.00 | 127.0 |
| 3.5 | 87.68 | 106.8 |
| 4.0 | 124.00 | 89.8 |
| 4.5 | 175.36 | 75.5 |
| 5.0 | 248.00 | 63.5 |
| 5.5 | 350.73 | 53.4 |
| 6.0 | 496.00 | 44.9 |
| 6.5 | 701.45 | 37.8 |
| 7.0 | 992.00 | 31.8 |

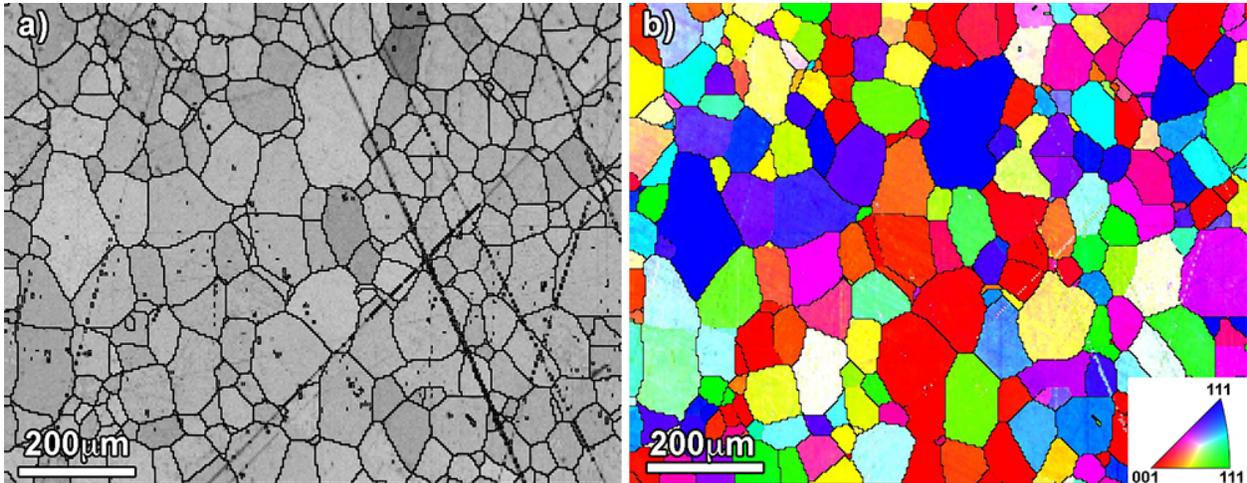

FIG. 3. (a) brightness and contrast image of kikuchi pattern intensity and (b) inverse pole figure of the A4 (TE1AES024) coupon, heat treated at 800 °C/3 hrs, from electron backscattered diffraction – orientation imaging microscopy (EBSD-OIM) with a 4 μm scanning step size. The legend describes color codes for grain orientation.

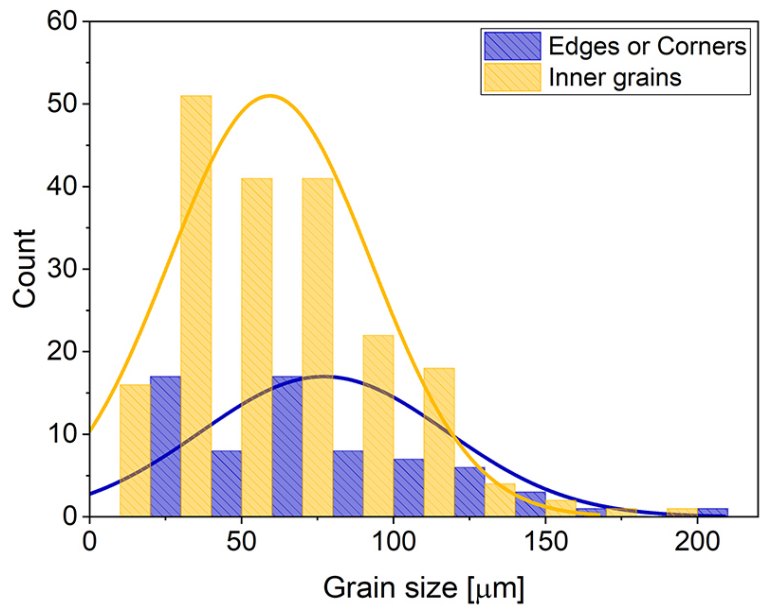

FIG. 4. Size distributions of inner (bright orange) and edges or corners (blue) grains, characterized with EBSD-OIM scanning on the A4 sample (TE1AES024) after 800 °C/3 hrs heat treatment (Fig. 3). Solid lines represent averaged histogram of frequency (counts).

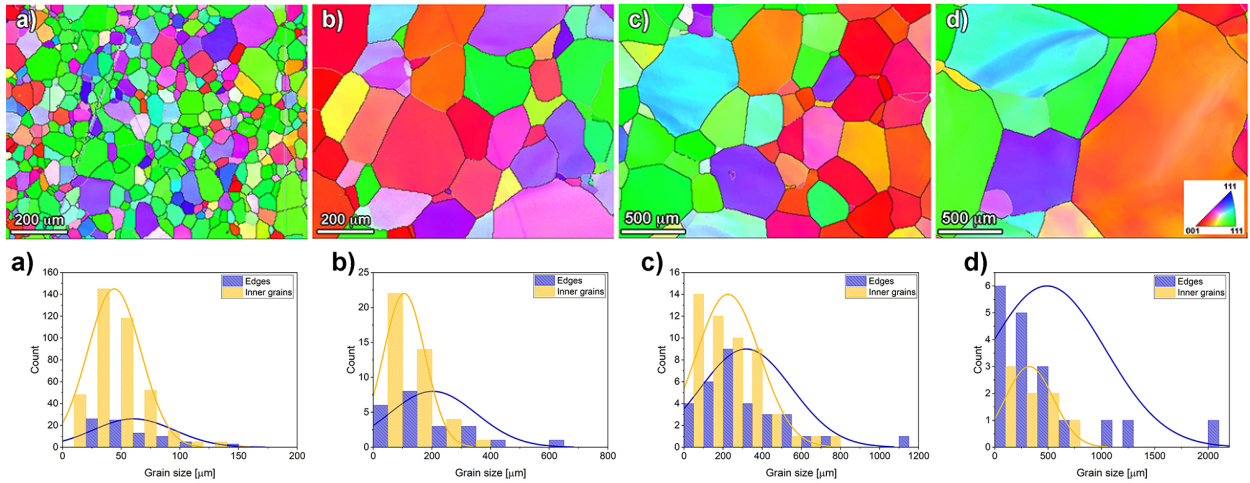

FIG 5. IPF (inverse pole figures) (top) and grain size distributions (bottom) of the S6 (SC06) coupon samples after thermally treated at (a) 800 °C, (b) 900 °C, (c) 950 °C, and (d) 1000 °C for 3 hours, evaluated by EBSD-OIM with 4 μm scan step size. Right orange and blue lines in the distribution profiles present averaged histogram of frequency (counts) of inner and edge or corner grains in the field of scan view, respectively.

Table 3. ASTM E112[13] Grain size No. ($G$) of the BCP'ed Nb samples analyzed by planimetric (Jeffries's) procedure.

| Cavity # | Coupon ID | 800 °C | 900 °C | 975 °C |
|---|---|---|---|---|
| TD02 | T2 | 5.5-6.0 | 1.5-2.0 | 1.0-1.5 |
| NX01 | N1 | 5.0-5.5 | 3.0-3.5 | 1.5-2.0 |
| NX02 | N2 | 7.0-7.5 | 6.0-6.5 | 4.5-5.0 |
| SC04 | S4 | 5.0-5.5 | 3.0-3.5 | 2.0-2.5 |
| SC06 | S6 | 6.0-6.5 | 2.0-2.5 | 0.5-1.0 |
| SC08 | S8 | 4.5-5.0 | 1.5-2.0 | 1.5-2.0 |
| SC02 | S2 | 4.0-4.5 | 1.0-1.5 | 0-1.0 |
| TE1AES024 | A4 | 4.0-4.5 | 1.5-2.0 | 1.0-1.5 |
| TE1AES022 | A2 | 5.5-6.0 | 3.0-3.5 | 1.5-2.0 |

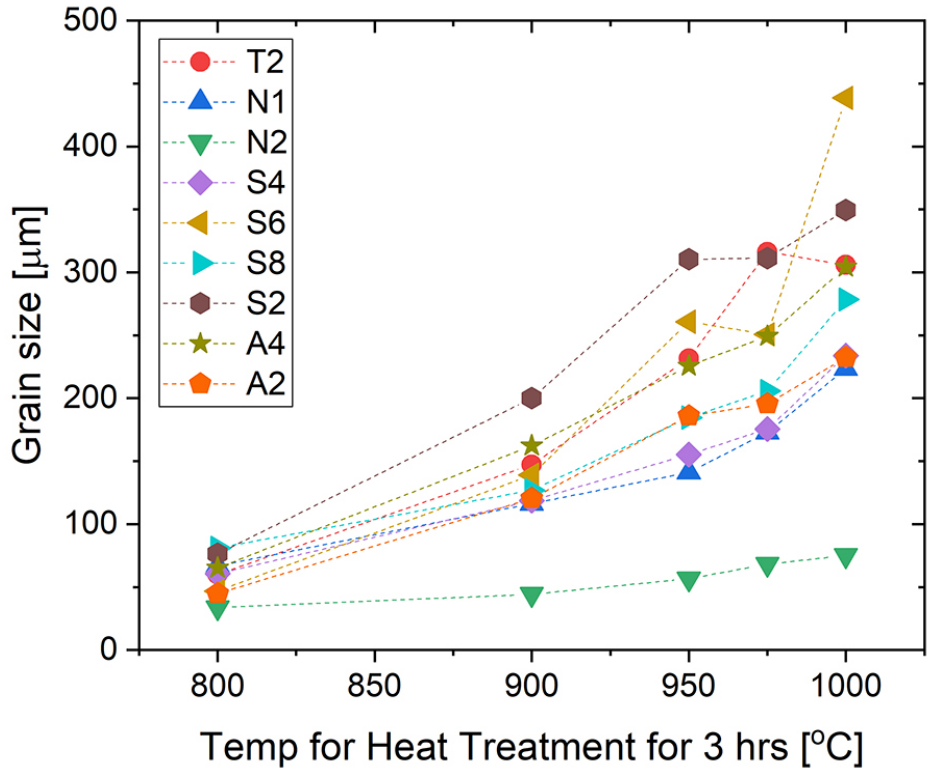

FIG. 6. Variation of average grain size of the coupon samples as a function of heat treatment temperature for 3 hours, from EBSD-OIM study.

Table 4. ASTM E112[13] Grain size No. (G) of the coupon samples based on grain diameters evaluated by EBSD-OIM study, with $B_{sc}/B_{nc}$ at $\Delta T = 5$ K.

| Cavity # | No | $B_{sc}/B_{nc}$ | 800 °C | 900 °C | 950 °C | 975 °C | 1000 °C |
|---|---|---|---|---|---|---|---|
| TD02 | T2 | 1.5838 | 5.0-5.5 | 2.5-3.0 | 1.0-1.5 | 0.5-1.0 | 0.5-1.0 |
| NX01 | N1 | 1.4378 | 5.0-5.5 | 3.4-4.0 | 2.5-3.0 | 2.0-2.5 | 1.5-2.0 |
| NX02[#] | N2 | 1.0842 | 6.5-7.0 | 6.0-6.5 | 5.0-5.5 | 4.5-5.0 | 4.5-5.0 |
| SC04 | S4 | 1.5948 | 5.0-5.5 | 3.5-4.0 | 2.0-2.5 | 2.0-2.5 | 1.0-1.5 |
| SC06 | S6 | 1.5726 | 5.5-6.0 | 3.0-3.5 | 1.0-1.5 | 1.0-1.5 | 00-0.0 |
| SC08 | S8 | 1.5618 | 4.0-4.5 | 3.0-3.5 | 1.5-2.0 | 1.5-2.0 | 0.5-1.0 |
| SC02 | S2 | 1.6851 | 4.5-5.0 | 2.0-2.5 | 0.5-1.0 | 0.5-1.0 | 0.0-0.5 |
| TE1AES024 | A4 | 1.6758 | 5.0-5.5 | 3.0-3.5 | 1.0-1.5 | 1.0-1.5 | 0.5-1.0 |
| TE1AES022 | A2 | 1.5206 | 6.0-6.5 | 3.0-3.5 | 1.5-2.0 | 1.5-2.0 | 1.0-1.5 |

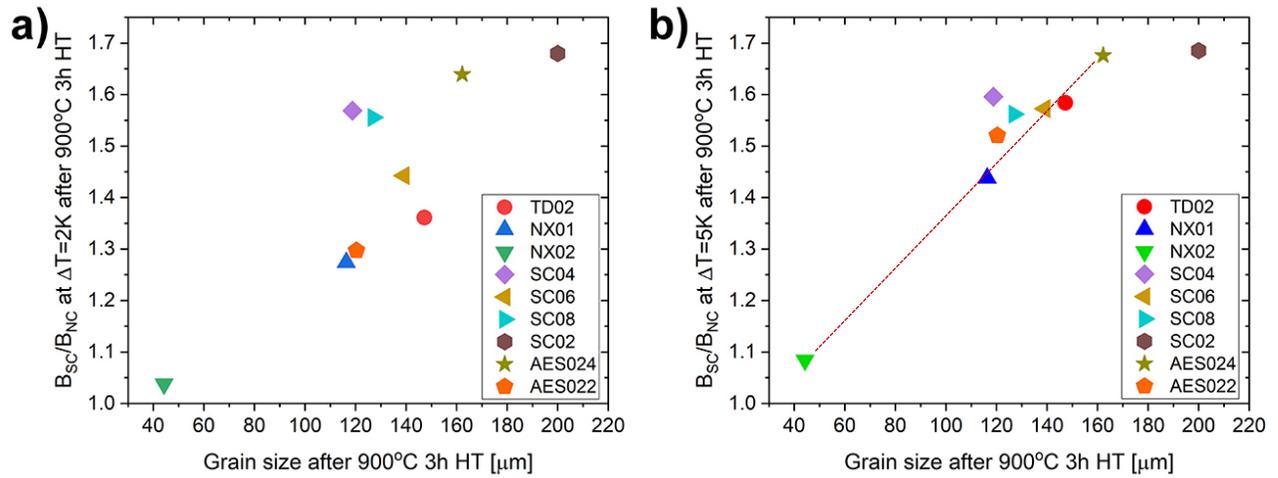

FIG. 7. Flux expulsion ratio ($B_{sc}/B_{nc}$) at $\Delta T = 2$ K (a) and 5 K (b) vs average grain sizes of the coupon samples after 900 °C/3 hrs HT, determined by EBSD-OIM analysis. The red dotted line is drawn for an eye guide of rough and monochromatic linear trend between $B_{sc}/B_{nc}$ and grain size.